\pdfoutput=1

\documentclass[11pt,twoside,a4paper,cmspaper,final,collab]{cms-tdr}
\def\svnVersion{437501P}\def\cmsCernNoTag{CERN-EP-2017-001}\def\cmsCernDate{\today}\def\cmsMessage{Published in Physics Letters B as \href{http://dx.doi.org/10.1016/j.physletb.2017.11.041}{\doi{10.1016/j.physletb.2017.11.041.}}}

\begin{document}\cmsNoteHeader{HIN-15-014}

\hyphenation{had-ron-i-za-tion}
\hyphenation{cal-or-i-me-ter}
\hyphenation{de-vices}
\RCS$Revision: 433416 $
\RCS$HeadURL: svn+ssh://svn.cern.ch/reps/tdr2/papers/HIN-15-014/trunk/HIN-15-014.tex $
\RCS$Id: HIN-15-014.tex 433416 2017-11-08 22:15:24Z alverson $

\newlength\cmsFigWidth
\ifthenelse{\boolean{cms@external}}{\setlength\cmsFigWidth{0.95\textwidth}}{\setlength\cmsFigWidth{0.95\textwidth}}

\newcommand{\sqrtsNN}{\ensuremath{\sqrt{\smash[b]{s_{_{\mathrm{NN}}}}}}\xspace}
\newcommand{\RAA}{\ensuremath{\mathrm{R_{AA}}}\xspace}
\newcommand{\AuAu}{\ensuremath{\mathrm{AuAu}}\xspace}
\newcommand{\PbPb}{\ensuremath{\mathrm{PbPb}}\xspace}
\newcommand{ \dmean }[1]{\langle\langle #1 \rangle\rangle}
\newcommand{ \cn }[1]{c_n\{#1\}}

\cmsNoteHeader{HIN-15-014}
\title{Azimuthal anisotropy of charged particles with transverse momentum up to 100\GeVc  in \PbPb  collisions at $\sqrtsNN=5.02\TeV$}

\date{\today}

\abstract{
The Fourier coefficients $v_2$ and $v_3$ characterizing the anisotropy of the azimuthal distribution of charged particles produced
in \PbPb  collisions at $\sqrtsNN= 5.02\TeV$ are measured with data collected by the CMS experiment.
The measurements cover a broad transverse momentum range, $1< \pt <100$\GeVc. The analysis focuses on the $\pt > 10$\GeVc  range,
where anisotropic azimuthal distributions should reflect the path-length dependence of parton energy loss in the created medium. Results are presented in several bins of \PbPb
collision centrality, spanning the 60\% most central events. The $v_2$ coefficient is measured with the scalar product and the multiparticle cumulant methods, which have different
sensitivities to initial-state fluctuations. The values from both methods remain positive up to $\pt \sim 60$--80\GeVc, in all examined centrality classes.
The $v_3$ coefficient, only measured with the scalar product method, tends to zero for $\pt \gtrsim 20$\GeVc. Comparisons between theoretical calculations and data provide
new constraints on the path-length dependence of parton energy loss in heavy ion collisions and highlight the importance of the initial-state fluctuations.
}

\hypersetup{%
pdfauthor={CMS Collaboration},%
pdftitle={Azimuthal anisotropy of charged particles with transverse momentum up to 100 GeV in PbPb collisions at sqrt(s[NN]) = 5.02 TeV},%
pdfsubject={CMS},%
pdfkeywords={CMS, QGP, high-pT, flow, parton energy loss, jet quenching}}

\maketitle

\section{Introduction}

Several observations made at RHIC in \AuAu  collisions at center-of-mass energy per nucleon pair
$\sqrtsNN = 200$\GeV~\cite{PHENIX, STAR, BRAHMS, PHOBOS}
and at the LHC in \PbPb  collisions at $\sqrtsNN = 2.76$ and
5.02\TeV~\cite{Aad:2010bu, Chatrchyan:2011sx, Aamodt:2010jd, CMS:2012aa, Aad:2015wga, Khachatryan:2016odn}
establish that high-energy partons
lose a significant fraction of their energy while traversing the hot and dense medium created in
these collisions. Measurements of the nuclear modification factor (\RAA),
a ratio that quantifies the modification of particle spectra between $\Pp\Pp$ and heavy ion collisions,
show a large suppression of high transverse-momentum (\pt) charged hadrons at RHIC~\cite{BRAHMSRaa,PHENIXRaa1,PHENIXRaa2,PHOBOSRaa,STARRaa1,STARRaa2}
and at LHC~\cite{Aamodt:2010jd,CMS:2012aa,Aad:2015wga,Khachatryan:2016odn}. Also, a strong asymmetry is observed in the energies of the two jets in dijet events
in \PbPb  collisions~\cite{Aad:2010bu,Chatrchyan:2011sx}.
These observations have triggered much progress in the understanding of jet
quenching phenomena, but do not provide sufficient information for a detailed
understanding of how the parton energy loss depends on the distance
traversed by the partons in the medium. The study of anisotropies in the
azimuthal angle distributions of high-\pt hadrons can
provide revealing information that is complementary to previous measurements.
These anisotropies are characterized by the $v_n$ coefficients of a Fourier expansion
in the distributions of azimuthal angle measured with respect to the event plane, defined
by the direction of maximum particle density in the transverse plane~\cite{Poskanzer:1998yz}.
Such studies have been performed at RHIC~\cite{Adare:2010sp} and at the
LHC~\cite{Chatrchyan:2012xq,Aad:2014vba,Adam:2016izf} up to $\pt \approx 10$ and 60\GeVc, respectively.
Most jet quenching models are unable to simultaneously reproduce
the \RAA  and $v_2$ measurements~\cite{MOLNAR2002495,Noronha:2010zc,Betz:2014cza}.
Nevertheless, recent attempts to solve
this puzzle have shown promise by considering initial-state collision geometry
asymmetries and fluctuations~\cite{Noronha-Hostler,Betz:2016ayq}, which
are predicted to strongly affect the high-\pt  $v_n$ coefficients, but not the \RAA values.
In particular, the fluctuations generate odd harmonics~\cite{Alver:2010gr} and
the measurement of the $v_3$ coefficient up to very
high \pt  is expected to clarify the importance of considering initial-state fluctuations
in the modeling of parton energy loss~\cite{Noronha-Hostler,Betz:2016ayq}.

In this Letter, the azimuthal anisotropy of charged particles produced in
\PbPb  collisions at $\sqrtsNN = 5.02\TeV$ is measured up to $\pt  \approx 100$\GeVc.
The scalar product (SP) method~\cite{Adler:2002pu,Luzum:2012da} is used to
determine the $v_{2}$ and $v_{3}$ coefficients as a function
of \pt  and collision centrality in the pseudorapidity range $\abs{\eta}<1$.
The unprecedented statistical reach of the \sqrtsNN = 5.02\TeV\ \PbPb  sample for high-\pt  particles allows
for the first precise measurement of the $v_{2}$ and $v_{3}$ coefficients at
high \pt. Furthermore, $v_{2}$ is also measured with the multiparticle
cumulant analysis method~\cite{Bilandzic:2010jr}, using 4-, 6- and 8-particle correlations.

\section{The CMS detector}

The central feature of the CMS apparatus is a superconducting solenoid of 6\unit{m} internal diameter providing a 3.8\unit{T} field. Within the solenoid volume there are a silicon pixel and strip tracker
detector, a lead tungstate crystal electromagnetic calorimeter (ECAL), and a brass and scintillator hadron calorimeter (HCAL), each composed of a barrel and two endcap sections. Muons are
measured in gas-ionization detectors embedded in the steel flux-return yoke outside the solenoid. The silicon tracker measures charged particles within $\abs{\eta} < 2.5$ and provides a \pt resolution
of about 1.5\% for 100\GeV charged particles.
Furthermore, the track impact parameter resolution is about 25--90 (45--150)\mum
in the transverse (longitudinal) dimension, depending on $\eta$ and \pt~\cite{Chatrchyan:2014fea}.
Iron and quartz-fiber Cherenkov hadron forward (HF) calorimeters cover the range
$2.9 < \abs{\eta} < 5.2$ on either side of the interaction region. The granularity
of the HF towers is $\Delta\eta{\times}\Delta\phi = 0.175{\times}0.175$ radians, allowing an
accurate reconstruction of the heavy ion event plane. A more detailed description of
the CMS detector, together with a definition of the coordinate system used and the
relevant kinematic variables, can be found in Ref.~\cite{bib_CMS}. The detailed Monte
Carlo simulation of the CMS detector response is based on \GEANTfour\ \cite{bib_geant}.

\section{Event and track selections}

The analysis of \PbPb  collisions is based on a data set corresponding to an integrated luminosity of 404\mubinv, collected in 2015.
Events were collected with
several trigger algorithms, composed of a hardware-based level 1 (L1) trigger,
followed by a software-based high-level trigger (HLT).
The \pt  region up to 14\GeVc  is covered by a minimum-bias trigger, which requires energy deposits in both HF calorimeters above a predefined threshold of approximately 1\GeV.
This minimum-bias trigger was prescaled during data taking.
To extend the measurement to higher order coefficients and higher \pt  (\eg, up to 100\GeVc),
a dedicated trigger that selects events containing a high-\pt  particle was used.
The L1 trigger requirement was based on the transverse energy (\ET) of the highest \ET  calorimeter region ($\Delta\eta{\times}\Delta\phi = 0.348{\times}0.348$)
in the barrel region ($\abs{\eta} < 1.044$). In the HLT farm, a fast version of the offline tracking algorithms was employed and the highest \pt  track was required to pass the strict selection
criteria described hereafter, resulting in a trigger efficiency of nearly 100\%. Different \ET and \pt  thresholds~\cite{Khachatryan:2016odn} were used at L1 and HLT, respectively,
to enrich the data sample with events that contain high-\pt  tracks.

In the offline analysis, an additional selection of hadronic collisions is applied by requiring at least
three towers with an energy deposit of more than 3\GeV per tower in each of the HF detectors.
The events are required to have a reconstructed primary vertex, formed by two or more tracks and required to have a distance from the nominal
interaction point of less than 15\unit{cm} along the beam axis and less than 0.15\unit{cm} in the transverse plane. The collision centrality in \PbPb events, i.e. the degree of overlap of the two colliding nuclei,
is determined from the \ET deposited in both HF calorimeters. Collision centrality bins are given in percentage ranges of the total hadronic cross section, 0--5\% corresponding
to the 5\% of collisions with the largest overlap of the two nuclei~\cite{Miller:2007ri}.

A standard CMS high-purity track selection~\cite{Chatrchyan:2014fea,Khachatryan:2015lha} is used to select primary tracks (tracks associated with the primary vertex). Additional requirements are
applied to enhance the purity of these primary tracks.
The track must be consistent with originating from the primary vertex by less than 3 standard deviations when estimating both the longitudinal and transverse distances of closest approach.
The relative uncertainty of the \pt  measurement, $\sigma(\pt)/\pt$, must be less than 10\%.
To ensure high tracking efficiency and reduce the rate of misreconstructed tracks, primary tracks are restricted to the
$\abs{\eta}<1$ and $\pt > 1$\GeVc region. Furthermore, tracks with $\pt  > 20$\GeVc
are required to match a compatible energy deposit in the calorimeters (ECAL + HCAL).
The tracking efficiency and detector acceptance in \PbPb  collisions are evaluated using simulated
{{\textsc{hydjet}} 1.9~\cite{Lokhtin:2005px} minimum bias and {\textsc{hydjet}}-embedded \PYTHIA~\cite{Sjostrand:2007gs} dijet events.
The combined geometrical acceptance and efficiency for primary track reconstruction, for $\pt > 1$\GeVc and $\abs{\eta}<1$, is 60--75\%, depending on centrality. Finally, the rate
of misreconstructed tracks reaches its
maximum in the most central events, where it approaches 10\%.

\section{Analysis technique}
The anisotropies of the particle azimuthal angle distributions are characterized by the $v_n$ Fourier coefficients, determined by the expansion
$\rd N/ \rd \phi \sim 1 + 2 \sum_n v_{n} \,\cos[n(\phi-\Psi_{n})]$, where $N$ is the number of particles and
$\Psi_{n}$ is the $n$th harmonic symmetry plane angle. Event-by-event variations in the initial energy density of the collision lead to the measured
event plane fluctuations about the (experimentally inaccessible) symmetry plane~\cite{Voloshin:1994mz}.
The SP method is used to measure azimuthal correlations and extract Fourier coefficients. In this method, the $v_n$ coefficients
can be expressed in terms of $Q_{n}$-vectors,
\begin{linenomath}
\begin{equation}
\label{eq:qsp}
{v_n}\left\{ {{\mathrm{SP}}} \right\} \equiv {\frac{\left\langle {{Q_n}Q_{nA}^*} \right\rangle }  {\sqrt {{\frac{\left\langle {Q_{nA}^{}Q_{nB}^*} \right\rangle \left\langle {Q_{nA}^{}Q_{nC}^*} \right\rangle } {\left\langle {Q_{nB}^{}Q_{nC}^*} \right\rangle }}} }},\text{ with }Q_{n},\, Q_{nA},\, Q_{nB},\, Q_{nC} \equiv \sum_{k=1}^M \omega_{k} \, \re^{in\phi_k},
\end{equation}
\end{linenomath}
where $M$ represents the number of tracks or HF towers with energy above
a certain threshold in each event, $\phi_k$ is the azimuthal angle of the $k^{\rm th}$ track or HF tower, and $\omega_k$ is a weighting factor equal to unity for $Q_n$, \pt for the tracks ($Q_{nC}$), and \ET
for the HF towers ($Q_{nA}$ and $Q_{nB}$). The angular brackets $\langle \rangle$ denote averages over all events. The $Q_n$ vector is based on the particles of interest, i.e., tracks with $\abs{\eta}<1$.
The $Q_{nA}$ and $Q_{nB}$ vectors are determined from the two HF calorimeters, covering the range $3<\abs{\eta}<5$, while the $Q_{nC}$ vector is
obtained using tracks with $\abs{\eta}<0.75$. If the particle of interest comes from the positive-$\eta$ side of the tracker, then $Q_{nA}$ is calculated using the negative-$\eta$ side of HF, and vice versa. The
large $\eta$ gap imposed between $Q_{nA}$ and $Q_{n}$ suppresses few-particle correlations, such as those induced by high-\pt\ jets and particle
decays, which do not depend on the event plane direction $\Psi^\mathrm{EP}_{n}$. The real part is taken for all averages of $Q$-vector products over the events. Azimuthal asymmetries that arise from the acceptance and other detector-related
effects are taken into account using a two-step process, where the $Q$-vector is first recentered and subsequently flattened~\cite{Barrette:1997pt}.
These corrections and their effects on the results are negligible for the CMS detector.
Since the measurements include correlations between low- and high-\pt  particles, the recently established event-plane decorrelation effect~\cite{Khachatryan:2015oea} cannot be neglected.
It is expected to reduce the $v_{n}$ values in comparison to those determined if the event
planes would be established exclusively using high-\pt  particles. The model calculations that include fluctuations in the initial state take into account this effect~\cite{Betz:2016ayq}.

The multiparticle cumulant method~\cite{Bilandzic:2010jr,Bilandzic:2013kga} is also used to measure $v_{2}$
from genuine 4-, 6-, and \mbox{8-particle} correlations, with the advantage of being less sensitive to few-particle
correlations, \eg, jet fragmentation. The cumulants are expressed in terms of the corresponding $Q_n$ vectors. We first define the 2-, 4-, 6-, and 8-particle correlators as

\begin{equation}
\begin{aligned}
\dmean{2} =& \left<\!\left< \re^{in(\phi_{1} -
  \phi_{2})} \right>\!\right>,\\
\dmean{4}=&
  \left<\!\left< \re^{in(\phi_{1} + \phi_{2} - \phi_{3}
  - \phi_{4})} \right>\!\right>,\\
\dmean{6}=&
  \left<\!\left< \re^{in(\phi_{1} + \phi_{2} + \phi_{3} - \phi_{4} - \phi_{5}
  - \phi_{6})} \right>\!\right>,\\
\dmean{8}=&
  \left<\!\left< \re^{in(\phi_{1} + \phi_{2} + \phi_{3} + \phi_{4} - \phi_{5} - \phi_{6} - \phi_{7}
  - \phi_{8})} \right>\!\right>,
\end{aligned}
\end{equation}
where the double average symbol $\dmean{~}$ indicates that the average is taken over all particle combinations and for all events.
The unbiased estimators of the reference multiparticle cumulants, $\cn$, are
defined as~\cite{Bilandzic:2013kga,Khachatryan:2015waa,Borghini:2001vi}
\ifthenelse{\boolean{cms@external}}{
\begin{equation}
\label{eq:cn}
\begin{aligned}
\cn{4} =& \dmean{4} - 2 \,  \dmean{2}^2,
\\
\cn{6} =& \dmean{6} - 9 \,  \dmean{4}\dmean{2} + 12 \,  \dmean{2}^3,
\\
\cn{8} =& \dmean{8} - 16 \,  \dmean{6}\dmean{2} -18 \,  \dmean{4}^2 \\
&+ 144 \,  \dmean{4}\dmean{2}^2 - 144 \,  \dmean{2}^{4}.
\end{aligned}
\end{equation}
}{
\begin{equation}
\label{eq:cn}
\begin{aligned}
\cn{4} =& \dmean{4} - 2 \,  \dmean{2}^2,
\\
\cn{6} =& \dmean{6} - 9 \,  \dmean{4}\dmean{2} + 12 \,  \dmean{2}^3,
\\
\cn{8} =& \dmean{8} - 16 \,  \dmean{6}\dmean{2} -18 \,  \dmean{4}^2 + 144 \,  \dmean{4}\dmean{2}^2 - 144 \,  \dmean{2}^4.
\end{aligned}
\end{equation}
}

In order to perform a measurement differential in \pt  in the multiparticle cumulant framework,
one of the particles in Eq.~(\ref{eq:cn}) is restricted to belong to a certain \pt bin.
Denoting by $\dmean{2'}$, etc., the modified particle correlators, the differential multiparticle cumulants are defined in Ref.~\cite{Borghini:2001vi} and can be derived as described in Ref. \cite{Bilandzic:2013kga},
\ifthenelse{\boolean{cms@external}}{
\begin{equation}
\label{eq:dndiff}
\begin{aligned}
d_{n}\{4\} =& \dmean{4'} - 2 \,  \dmean{2'}\dmean{2},\\
d_{n}\{6\} =& \dmean{6'} - 6 \,  \dmean{4'}\dmean{2} - 3 \,  \dmean{2'}\dmean{4} \\
&+ 12 \,  \dmean{2'} \dmean{2}^2,\\
d_{n}\{8\} =& \dmean{8'} - 12 \,  \dmean{6'}\dmean{2} - 4 \,  \dmean{2'}\dmean{6} \\
&- 18 \,  \dmean{4'}\dmean{4}
             +\, 72 \,  \dmean{4'}\dmean{2}^2 \\
             &+ 72 \,  \dmean{4}\dmean{2}\dmean{2'}  - 144 \,  \dmean{2'}\dmean{2}^3.
\end{aligned}
\end{equation}
}{
\begin{equation}
\label{eq:dndiff}
\begin{aligned}
d_{n}\{4\} =& \dmean{4'} - 2 \,  \dmean{2'}\dmean{2},\\
d_{n}\{6\} =& \dmean{6'} - 6 \,  \dmean{4'}\dmean{2} - 3 \,  \dmean{2'}\dmean{4} + 12 \,  \dmean{2'} \dmean{2}^2,\\
d_{n}\{8\} =& \dmean{8'} - 12 \,  \dmean{6'}\dmean{2} - 4 \,  \dmean{2'}\dmean{6} - 18 \,  \dmean{4'}\dmean{4}  \\
            & +\, 72 \,  \dmean{4'}\dmean{2}^2 + 72 \,  \dmean{4}\dmean{2}\dmean{2'} - 144 \,  \dmean{2'}\dmean{2}^3.
\end{aligned}
\end{equation}
}

Finally, with respect to the reference multiparticle cumulants, the differential 4-, 6-, and \mbox{8-particle} $v_{n}(\pt, \eta)$ coefficients are derived as
\begin{equation}
\begin{aligned}
v_{n}\{4\}(\pt, \eta) =& - d_n\{4\} \,  (-c_n\{4\})^{-3/4}, \\
v_{n}\{6\}(\pt, \eta) =& \hphantom{-}d_n\{6\} \,  (c_n\{6\})^{-5/6}  \,  4^{-1/6} ,\\
v_{n}\{8\}(\pt, \eta) =& - d_n\{8\} \,  (-c_n\{8\})^{-7/8} \,  33^{-1/8}.
\end{aligned}
\end{equation}

The statistical uncertainties are evaluated with a data-driven method, as previously employed in Ref.~\cite{Khachatryan:2015waa}. The data set is divided into 10 subsets with
roughly equal numbers of events and the standard deviation of the resulting distribution of the cumulant is used to estimate the uncertainties.

\section{Systematic uncertainties}

At low \pt, the relative systematic uncertainties for $v_{2}\{\mathrm{SP}\}$ and $v_{3}\{\mathrm{SP}\}$ are found to be similar.
At high \pt, the $v_{3}\{\mathrm{SP}\}$ statistical uncertainties are too large to properly disentangle statistical fluctuations
from systematic effects. Therefore, the $v_{2}$ systematic uncertainties, expressed in terms of relative values in \%, are applied to
$v_{3}$, with the exception of the uncertainties due to the few-particle
correlations, discussed below. The systematic uncertainties due to the vertex position
selection and to the \pt  dependence of the tracking efficiency corrections are common to the SP and cumulant analyses.
They are found to be less than 1\% and independent of \pt and centrality.
The systematic uncertainties due to misreconstructed tracks are derived by changing the track selection criteria.
The results are found to depend on \pt  but not centrality, and are also different for the cumulant and SP methods.
The track selection uncertainties have been found to gradually increase from
$\sim$2\% at low \pt  to $\sim$50\% for $\pt> 60$\GeVc for the SP method, and from $\sim$2\% to $\sim$12\% for the cumulant analysis.
The SP results have an additional uncertainty arising from few-particle correlations. This uncertainty is determined by varying the $\eta$ gap and
contributes differently to the $v_{2}$ and $v_{3}$ measurements. It is found to depend on both \pt  and centrality, and ranges in absolute value from 0 to 0.022 for $v_{2}$ and
from 0 to 0.030 for $v_3$.

\section{Results}

Figure~\ref{fig:v2v3} shows the $v_{2}$ and $v_{3}$ results obtained from the
SP method as a function of \pt, up to about 100\GeVc, in seven collision centrality ranges. From low- to high-\pt, the $v_2$ and $v_3$
values first increase with increasing \pt, up to a maximum near $\pt \approx  3$\GeVc, before decreasing again.
In most centrality ranges, $v_2$ remains positive up to $\pt \sim 60$--80\GeVc, becoming consistent with zero at higher \pt.
Positive $v_3$ values are found up to $\pt \approx 20$\GeVc  over the 0--40\% centrality range. At higher \pt, the measured $v_3$ value is
consistent with zero within the experimental uncertainties. Given the systematic uncertainties, the measured values are compatible with zero.
Some negative $v_{3}$ values are seen at high \pt  in the 40--50\% centrality range, but such peripheral events are the most contaminated by
back-to-back jet correlations. This is confirmed by studying the $\eta$ gap dependence of the results in both measured and simulated events,
where the latter include dijets embedded into {\textsc{hydjet}} events with zero input anisotropy.
In the centrality range 50--60\%, $v_{3}$ is only measured up to 20\GeVc
because of lack of events containing higher \pt  particles.

\ifthenelse{\boolean{cms@external}}{
\begin{figure*}[htb]
        \centering
	\includegraphics[width=\cmsFigWidth]{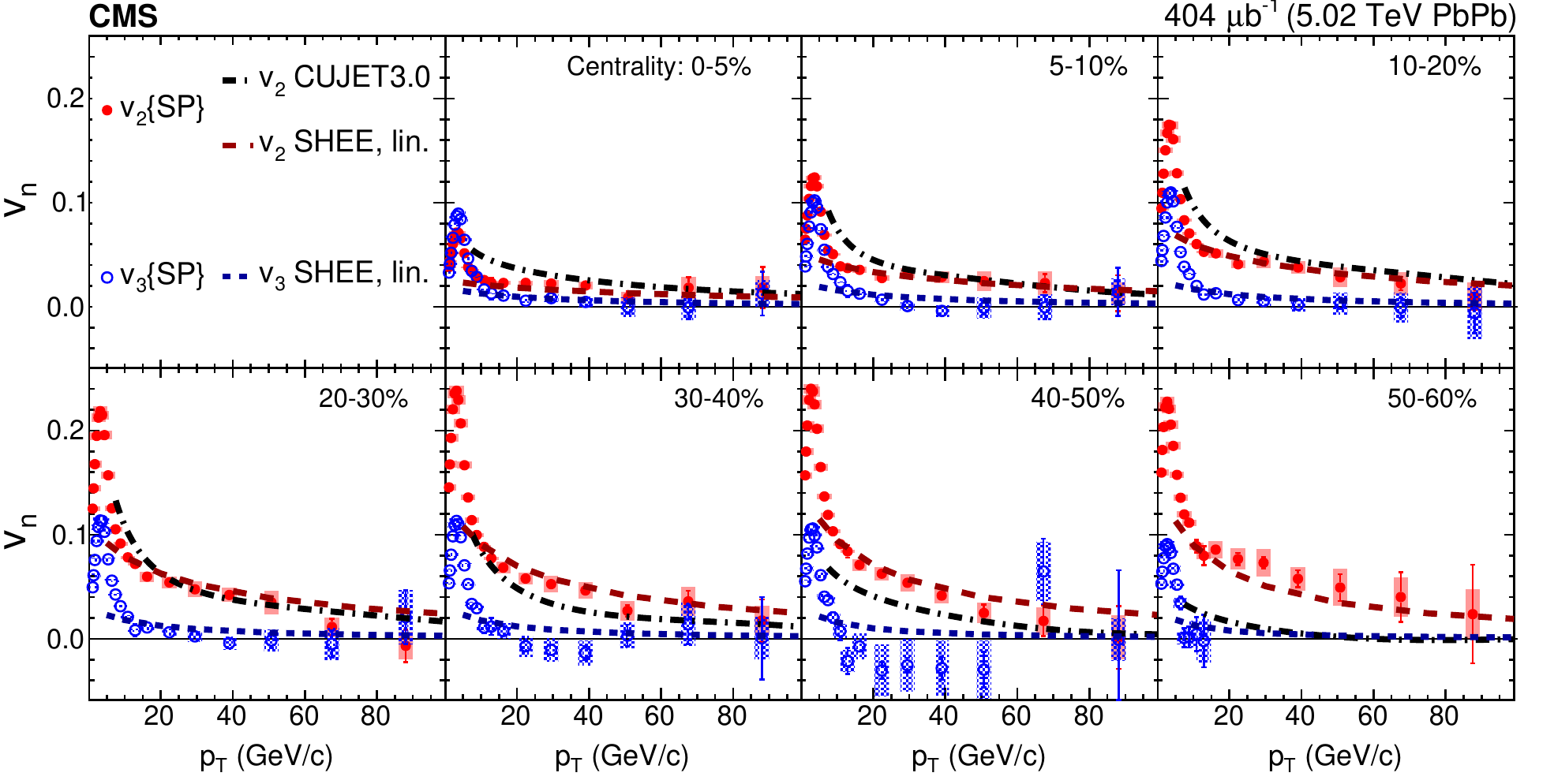}
	\caption{The $v_2$ and $v_3$ results from the SP method as a function of \pt,
	    in seven collision centrality ranges from 0--5\%
	    to 50--60\%. The vertical bars (shaded boxes)
        represent the statistical (systematic) uncertainties.
        The curves represent calculations made with the CUJET3.0~\cite{Xu:2015bbz} and the SHEE models~\cite{Betz:2016ayq} (see text).}
	\label{fig:v2v3}
\end{figure*}
}{
\begin{figure*}[htbp]
        \centering
	\includegraphics[width=0.9\textwidth]{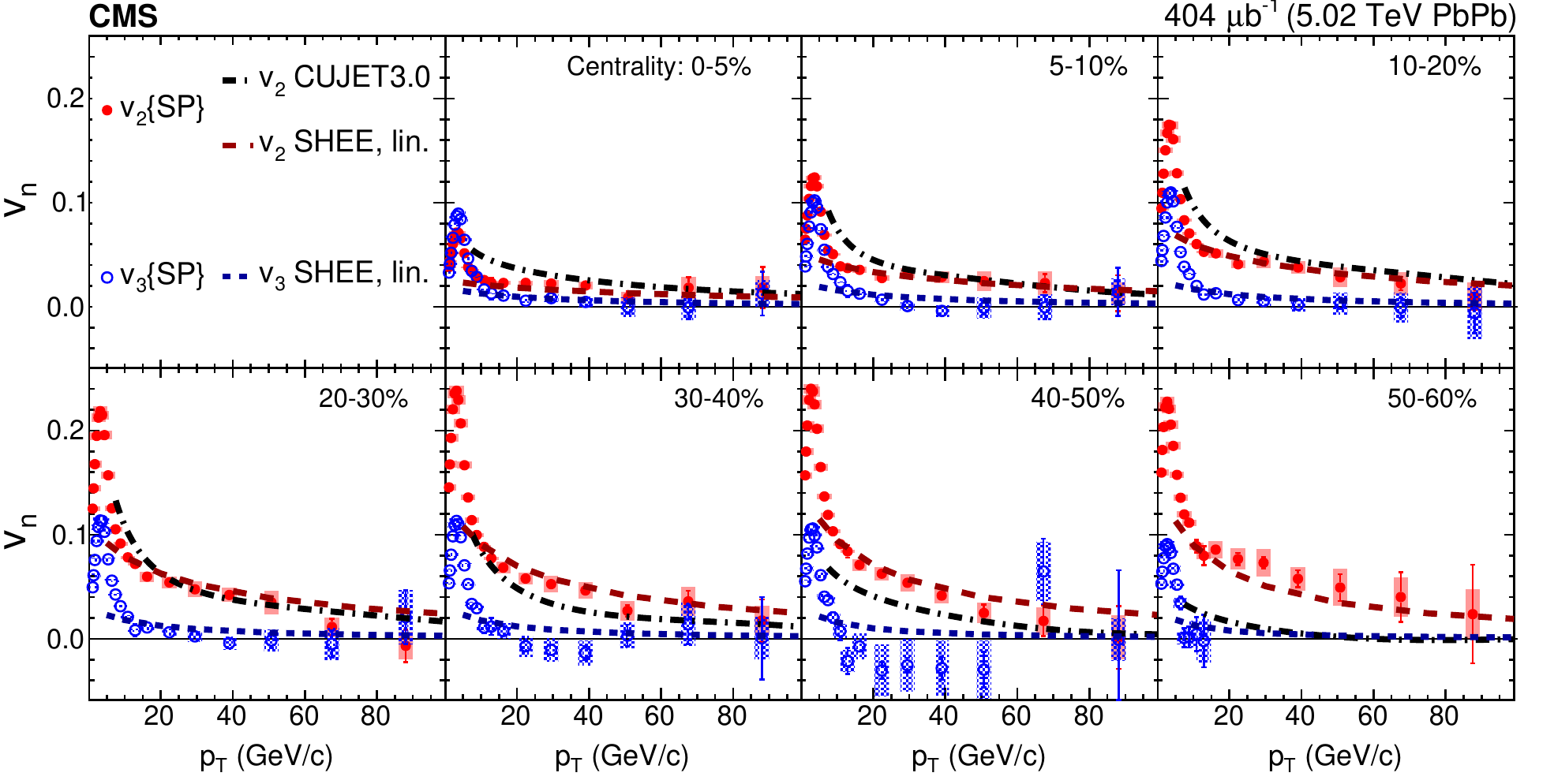}
	\caption{The $v_2$ and $v_3$ results from the SP method as a function of \pt,
	    in seven collision centrality ranges from 0--5\%
	    to 50--60\%. The vertical bars (shaded boxes)
        represent the statistical (systematic) uncertainties.
        The curves represent calculations made with the CUJET3.0~\cite{Xu:2015bbz} and the SHEE models~\cite{Betz:2016ayq} (see text).}
	\label{fig:v2v3}
\end{figure*}
}

The $v_{2}$ and $v_3$ results are compared to the CUJET3.0~\cite{Xu:2015bbz} and
SHEE~\cite{Noronha-Hostler} models for several centrality bins. A key difference
between these two models is that the SHEE framework includes initial-state geometry
fluctuations, while CUJET3.0 uses a smooth hydrodynamic background. The CUJET3.0 model
uses perturbative quantum chromodynamics (pQCD) calculations to describe
the hard parton interactions in the quark-gluon plasma (QGP), complemented
by a perfect-fluid hydrodynamic expansion of the medium. The SHEE calculations use viscous hydrodynamics including
event-by-event fluctuations in the soft sector~\cite{Betz:2016ayq,Noronha-Hostler:2013gga,Noronha-Hostler:2014dqa},
in addition to an energy loss model~\cite{Betz:2016ayq,Betz:2011tu,Betz:2012qq}. They are performed with
a low shear viscosity to entropy density ratio ($\eta/s$), less than or equal to 0.12
(although higher values do not affect the high-\pt  predictions),
a chemical freezout temperature of 160\MeV, and a linear path-length
dependence of the energy loss inspired by pQCD, similar to that in CUJET3.0. In addition, both model calculations are only valid for $\pt > 10$\GeVc.

Over the full centrality range, the CUJET3.0 calculations describe qualitatively the trend observed in the $v_{2}$ data for \pt $> 10$\GeVc,
but fail to quantitatively reproduce the results. For instance,
in the centrality range 0--30\% and for $10< \pt <40$\GeVc, $v_2$ is overestimated by 10--50\%, while the model largely underestimates it in the peripheral bins. The SHEE calculations of both $v_2$ and $v_3$
are in good agreement with the data for $\pt  > 10$\GeVc  over the full centrality range.
The success of the SHEE framework suggests that modeling the initial-state fluctuations may be a crucial ingredient to describe the experimental data related to parton energy loss.
Although not shown in the figure, a scenario in the SHEE framework with a quadratic path-length
dependence of the energy loss, inspired by gauge-gravity duality~\cite{Gubser:2008as,Dominguez:2008vd},
was also tested and seen to disagree with the data. As just one example, this alternative path-length dependence is found to
overestimate the data by 30--40\% for $\pt > 20$\GeVc  in the 20--30\% centrality range.

The $v_{2}$ values are also obtained from 4-, 6-, and 8-particle cumulant analyses, as shown in Fig.~\ref{fig:vn_all}, where the SP $v_2$ results are also included for comparison.
For $\pt< 3$\GeVc, the results follow the expectation from Bessel-Gaussian or elliptic power $v_2$ distributions, which predict
$v_{2}\{\mathrm{SP}\} > v_{2}\{4\} \approx v_{2}\{6\} \approx v_{2}\{8\}$~\cite{Yan:2014nsa,Voloshin:2007pc,Chatrchyan2013213}.
The observation that the multiparticle cumulant values remain similar up to \pt  $= 100$\GeVc  ($v_{2}\{4\} \approx v_{2}\{6\} \approx v_{2}\{8\}$),
further suggests that the azimuthal anisotropy is strongly affected by the initial-state geometry and its event-by-event fluctuations~\cite{Betz:2016ayq,Noronha-Hostler}.
At higher \pt, the difference between SP and multiparticle cumulant results shows a tendency to decrease. Nevertheless, the uncertainities are too large to draw a firm conclusion.
This tendency might be due to \pt dependence of flow vector fluctuations, which depends on the shear viscosity over entropy density ratio of the medium~\cite{Betz:2016ayq,Acharya:2017ino}.
Therefore, the results presented in Fig.~\ref{fig:vn_all} provide important information to constrain the QGP shear viscosity
in \PbPb\ collisions.

\ifthenelse{\boolean{cms@external}}{
\begin{figure*}[htb]
        \centering
	\includegraphics[width=\cmsFigWidth]{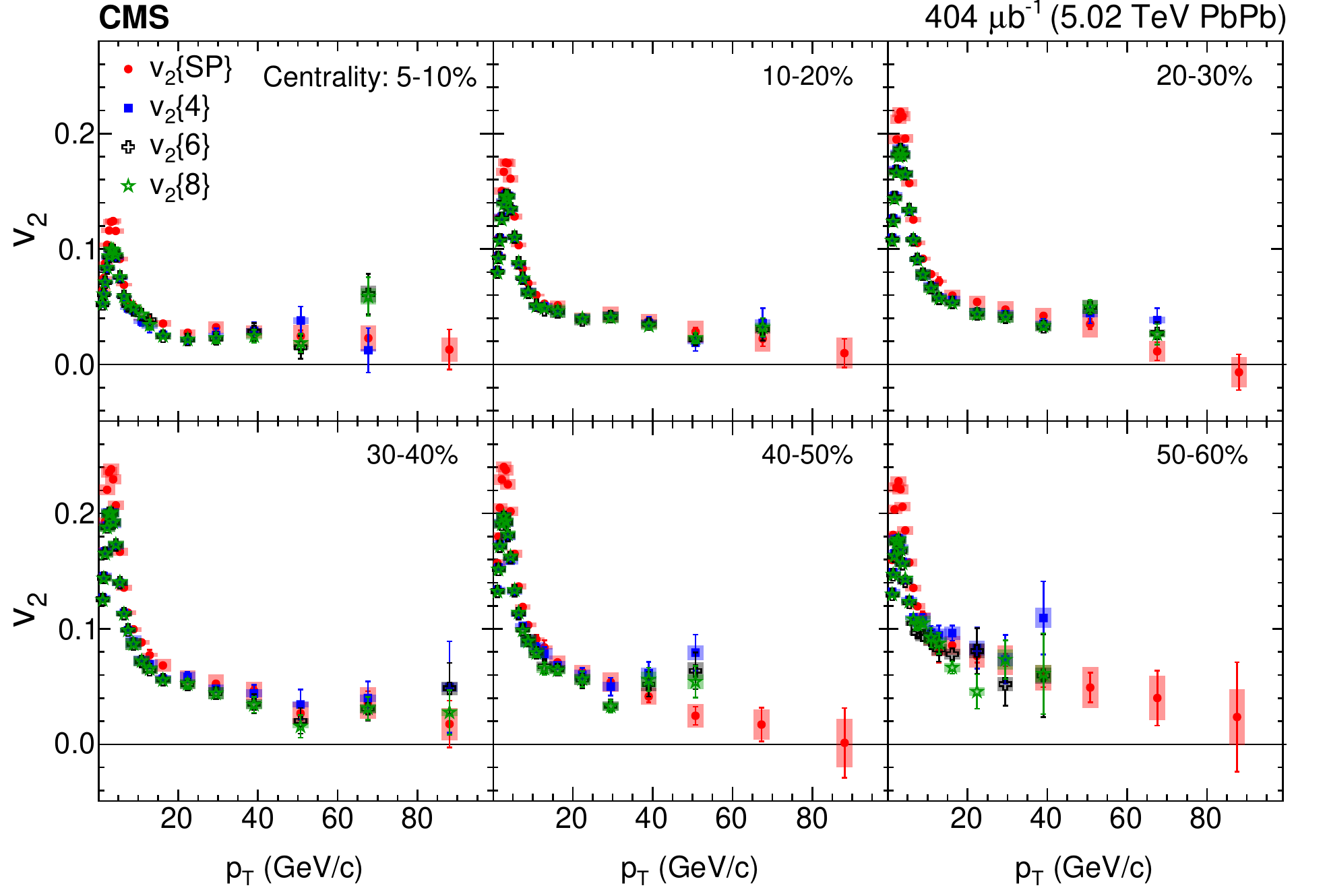}
	\caption{
	Comparison between the $v_2$ results from the SP
	and the 4-, 6-, and 8-particle
        cumulant methods, as a function of \pt, in six centrality ranges
	from 0--5\% to 50--60\%.
	The vertical bars (shaded boxes) represent the statistical (systematic) uncertainties.}
	\label{fig:vn_all}
\end{figure*}
}{
\begin{figure*}[htbp]
        \centering
	\includegraphics[width=\cmsFigWidth]{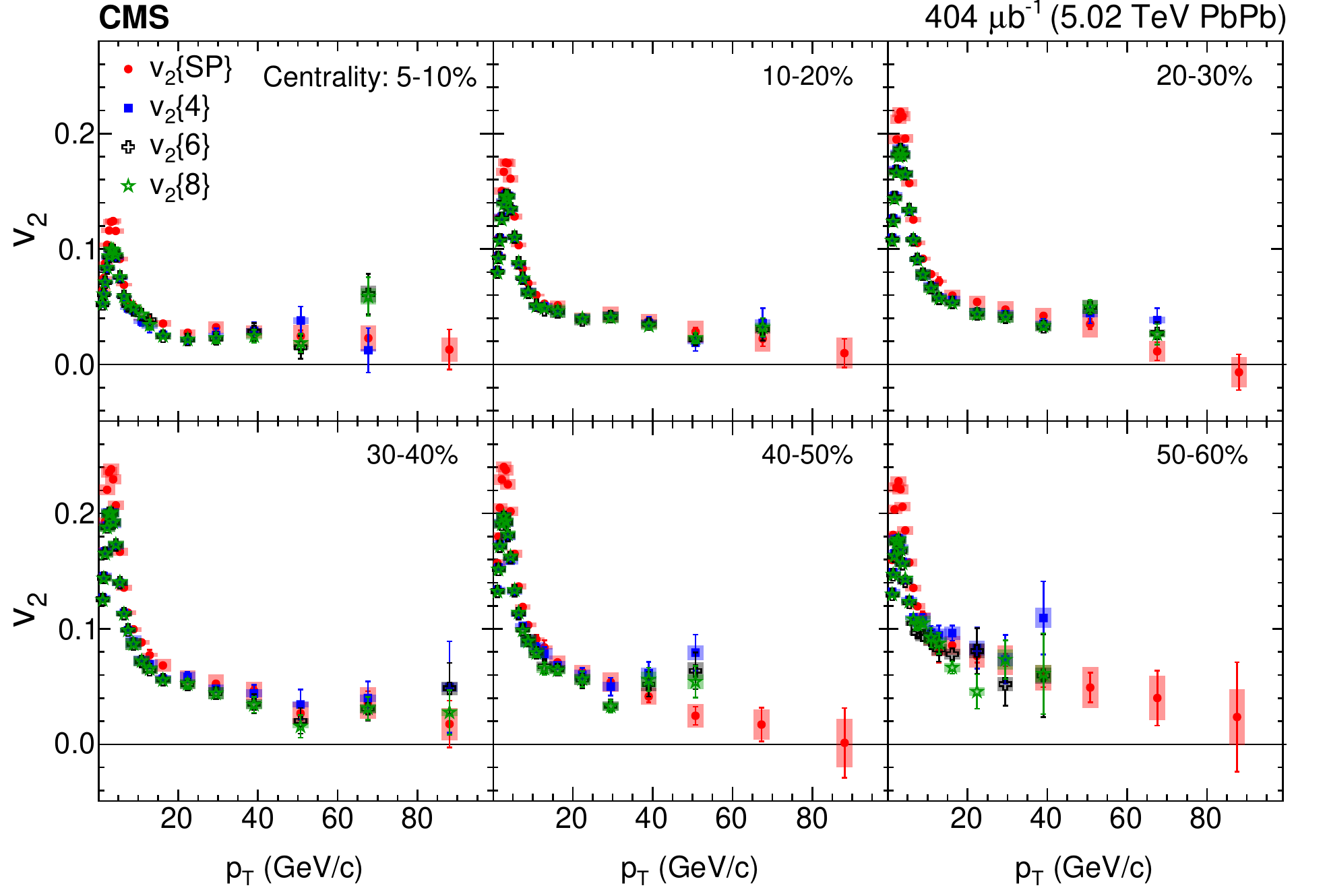}
	\caption{
	Comparison between the $v_2$ results from the SP
	and the 4-, 6-, and 8-particle
        cumulant methods, as a function of \pt, in six centrality ranges
	from 0--5\% to 50--60\%.
	The vertical bars (shaded boxes) represent the statistical (systematic) uncertainties.}
	\label{fig:vn_all}
\end{figure*}
}
Figure~\ref{fig:v2HiLow} shows the correlation between high-\pt  and low-\pt  $v_{2}$ values, for investigating the connection between the azimuthal anisotropies induced by hydrodynamic
flow and the path-length dependence of parton energy loss~\cite{Betz:2016ayq,Noronha-Hostler}.
The most peripheral $v_{2}\{\mathrm{SP}\}$ and $v_{2}\{4\}$ data points are the ones with the largest error bars.
Linear fits to the centrality dependent $v_2$ correlation between the low- and high-\pt regions are shown in the figure.
Here a zero intercept is assumed. The corresponding $\chi^{2}$ over the number of degree of freedom values are found to be near 1--1.5,
except for the $26< \pt <35$\GeVc range, where a positive intercept is indicated for the $v_{2}\{\mathrm{SP}\}$ results.
The non-zero intercept might reflect a centrality dependent event-plane decorrelation that increases going to higher \pt.
The slope values for $v_{2}\{\mathrm{SP}\}$ and $v_{2}\{4\}$ are found to be compatible within statistical uncertainties and to decrease when selecting higher \pt particles.
This suggests that the initial-state geometry and its fluctuations are likely to be the common causes of the observed particle azimuthal anisotropies at both low and high \pt.

\begin{figure*}[tbh]
	\includegraphics[width=\linewidth]{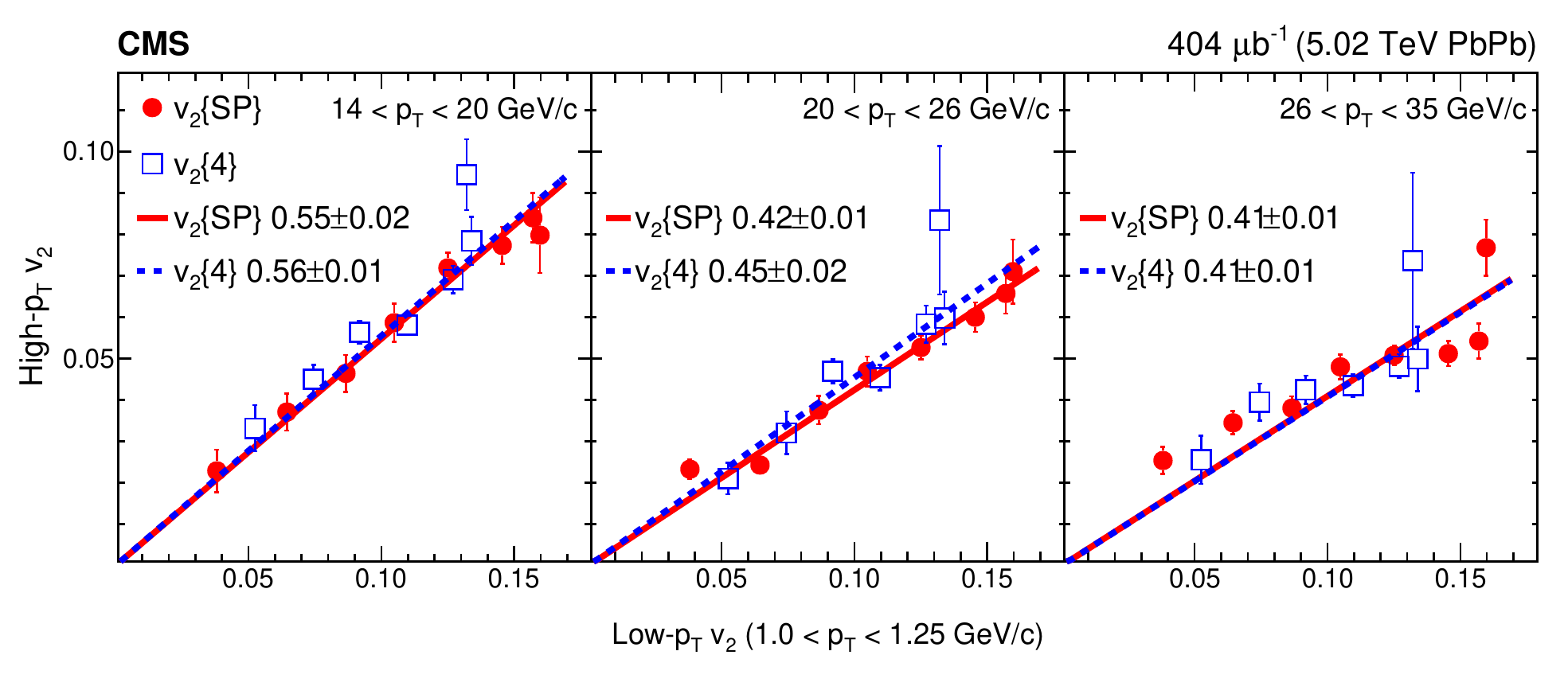}
	\caption{Correlation between the high-\pt  $v_{2}$ measured in the 14--20 (left), 20--26 (middle), and 26--35\GeVc (right)
                 \pt  ranges and the low-\pt  $v_{2}$ measured in the $1<\pt<1.25$\GeVc range, with the SP (closed circles) and cumulant (open squares)
                 methods. The points represent the centrality bins 0--5, 5--10, 10--15, 15--20, 20--30, 30--40, 40--50, and 50--60\% for the SP results.
                 For the cumulant method, the bin 0--5\% is not shown. Lines represent a linear fit to the SP results (red) and cumulant results (dashed blue).}
	\label{fig:v2HiLow}
\end{figure*}

\section{Summary}

The azimuthal anisotropy of charged particles produced in \PbPb collisions at $\sqrtsNN = 5.02$\TeV
has been studied using data collected by the CMS experiment. The $v_2$ and $v_3$ coefficients are
determined, as a function of collision centrality, over the widest transverse momentum range studied to date
(from 1 up to 100\GeVc). For the first time, the multiparticle cumulant method is used for $\pt > 20$\GeVc.
Over the measured centrality range, positive $v_2$ values are found up to $\pt \sim 60$--80\GeVc,
while the $v_3$ values are consistent with zero for \pt  $> 20$\GeVc.
For $\pt < 3$\GeVc, $v_{2}\{\mathrm{SP}\} > v_{2}\{4\} \approx v_{2}\{6\} \approx v_{2}\{8\}$,
consistent with a collective behavior arising from the hydrodynamic expansion of a quark-gluon plasma.
The similarity of $v_2\{\mathrm{SP}\}$, $v_{2}\{4\}$, $v_{2}\{6\}$, and $v_{2}\{8\}$ at high \pt  suggests that $v_{2}$ originates from the path-length dependence of parton energy loss associated
with an asymmetric initial collision geometry. In addition, a common trend in the centrality dependence of $v_{2}$ is
observed over the full \pt range, further supporting a common connection to the initial-state geometry and its fluctuations.
A model calculation (SHEE) incorporating initial-state fluctuations with a linear
path-length dependence of parton energy loss is found to be in good agreement with the data,
over the wide \pt and centrality ranges probed in this analysis.

\begin{acknowledgments}
We congratulate our colleagues in the CERN accelerator departments for the excellent performance of the LHC and thank the technical and administrative staffs at CERN and at other CMS institutes for their contributions to the success of the CMS effort. In addition, we gratefully acknowledge the computing centers and personnel of the Worldwide LHC Computing Grid for delivering so effectively the computing infrastructure essential to our analyses. Finally, we acknowledge the enduring support for the construction and operation of the LHC and the CMS detector provided by the following funding agencies: BMWFW and FWF (Austria); FNRS and FWO (Belgium); CNPq, CAPES, FAPERJ, and FAPESP (Brazil); MES (Bulgaria); CERN; CAS, MoST, and NSFC (China); COLCIENCIAS (Colombia); MSES and CSF (Croatia); RPF (Cyprus); SENESCYT (Ecuador); MoER, ERC IUT, and ERDF (Estonia); Academy of Finland, MEC, and HIP (Finland); CEA and CNRS/IN2P3 (France); BMBF, DFG, and HGF (Germany); GSRT (Greece); OTKA and NIH (Hungary); DAE and DST (India); IPM (Iran); SFI (Ireland); INFN (Italy); MSIP and NRF (Republic of Korea); LAS (Lithuania); MOE and UM (Malaysia); BUAP, CINVESTAV, CONACYT, LNS, SEP, and UASLP-FAI (Mexico); MBIE (New Zealand); PAEC (Pakistan); MSHE and NSC (Poland); FCT (Portugal); JINR (Dubna); MON, RosAtom, RAS, RFBR and RAEP (Russia); MESTD (Serbia); SEIDI, CPAN, PCTI and FEDER (Spain); Swiss Funding Agencies (Switzerland); MST (Taipei); ThEPCenter, IPST, STAR, and NSTDA (Thailand); TUBITAK and TAEK (Turkey); NASU and SFFR (Ukraine); STFC (United Kingdom); DOE and NSF (USA).

\hyphenation{Rachada-pisek} Individuals have received support from the Marie-Curie program and the European Research Council and EPLANET (European Union); the Leventis Foundation; the A. P. Sloan Foundation; the Alexander von Humboldt Foundation; the Belgian Federal Science Policy Office; the Fonds pour la Formation \`a la Recherche dans l'Industrie et dans l'Agriculture (FRIA-Belgium); the Agentschap voor Innovatie door Wetenschap en Technologie (IWT-Belgium); the Ministry of Education, Youth and Sports (MEYS) of the Czech Republic; the Council of Science and Industrial Research, India; the HOMING PLUS program of the Foundation for Polish Science, cofinanced from European Union, Regional Development Fund, the Mobility Plus program of the Ministry of Science and Higher Education, the National Science Center (Poland), contracts Harmonia 2014/14/M/ST2/00428, Opus 2014/13/B/ST2/02543, 2014/15/B/ST2/03998, and 2015/19/B/ST2/02861, Sonata-bis 2012/07/E/ST2/01406; the National Priorities Research Program by Qatar National Research Fund; the Programa Clar\'in-COFUND del Principado de Asturias; the Thalis and Aristeia programs cofinanced by EU-ESF and the Greek NSRF; the Rachadapisek Sompot Fund for Postdoctoral Fellowship, Chulalongkorn University and the Chulalongkorn Academic into Its 2nd Century Project Advancement Project (Thailand); and the Welch Foundation, contract C-1845. \end{acknowledgments}

\bibliography{auto_generated}

\cleardoublepage \appendix\section{The CMS Collaboration \label{app:collab}}\begin{sloppypar}\hyphenpenalty=5000\widowpenalty=500\clubpenalty=5000\textbf{Yerevan Physics Institute,  Yerevan,  Armenia}\\*[0pt]
A.M.~Sirunyan, A.~Tumasyan
\vskip\cmsinstskip
\textbf{Institut f\"{u}r Hochenergiephysik,  Wien,  Austria}\\*[0pt]
W.~Adam, E.~Asilar, T.~Bergauer, J.~Brandstetter, E.~Brondolin, M.~Dragicevic, J.~Er\"{o}, M.~Flechl, M.~Friedl, R.~Fr\"{u}hwirth\cmsAuthorMark{1}, V.M.~Ghete, C.~Hartl, N.~H\"{o}rmann, J.~Hrubec, M.~Jeitler\cmsAuthorMark{1}, A.~K\"{o}nig, I.~Kr\"{a}tschmer, D.~Liko, T.~Matsushita, I.~Mikulec, D.~Rabady, N.~Rad, B.~Rahbaran, H.~Rohringer, J.~Schieck\cmsAuthorMark{1}, J.~Strauss, W.~Waltenberger, C.-E.~Wulz\cmsAuthorMark{1}
% ...
\vskip\cmsinstskip
% ...
1:~~Also at Vienna University of Technology, Vienna, Austria\\
% ...

\end{sloppypar}
\end{document}